\documentclass{PoS}

\usepackage{bm}
\usepackage{color,xcolor}
\usepackage{mathtools,amsbsy}
\usepackage{keyval,graphicx}
\usepackage{slashed}
\usepackage{epstopdf}
\usepackage{isotope}
\usepackage{dcolumn}

\title{Exotic candidates with heavy quark(s)}

\ShortTitle{Exotic candidates with heavy quark(s)}

\author{\speaker{Qian Wang}\\
        Helmholtz-Institut f\"ur Strahlen-und Kernphysik and Bethe Center for Theoretical Physics, Universit\"at Bonn\\
        E-mail: \email{wangqian@hiskp.uni-bonn.de}}

\abstract{In the last decades, the number of exotic candidates which are beyond the conventional quark model has grown dramatically. 
At the same time, numerous theoretical interpretations, such as tetraquark, hybrid, 
hadroquarkonium and hadronic molecule, have been proposed to understand 
their nature. In principle, all the configurations with the same quantum numbers
 can mix with each other.  
Thus at present we aim at identifying the prominent component. 
 As an extended object, a hadronic 
 molecule has some distinguishable features from other more compact configurations.
In the end, the two $Z_b$ states are used to illustrate some features of the molecular scenario.}

\FullConference{XVII International Conference on Hadron Spectroscopy and Structure - Hadron2017\\
		25-29 September, 2017\\
		University of Salamanca, Salamanca, Spain}

\begin{document}

\section{Introduction}
\begin{figure}
\begin{center}
\includegraphics[width=0.25\textwidth]{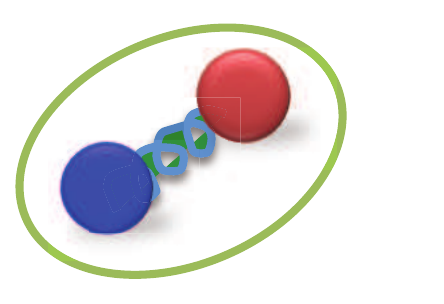}\hspace{0.1cm} \includegraphics[width=0.22\textwidth]{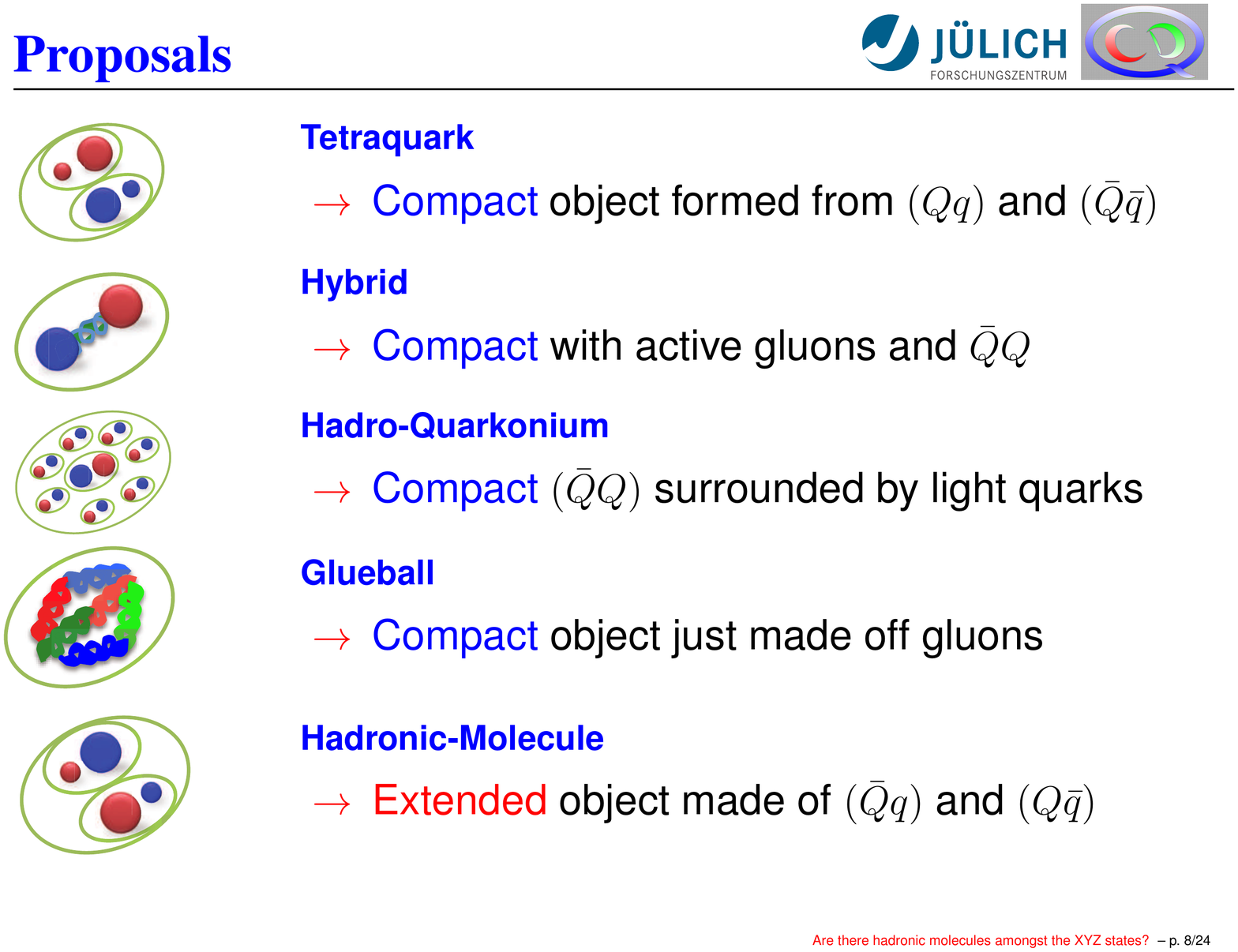}\hspace{0.4cm} \includegraphics[width=0.19\textwidth]{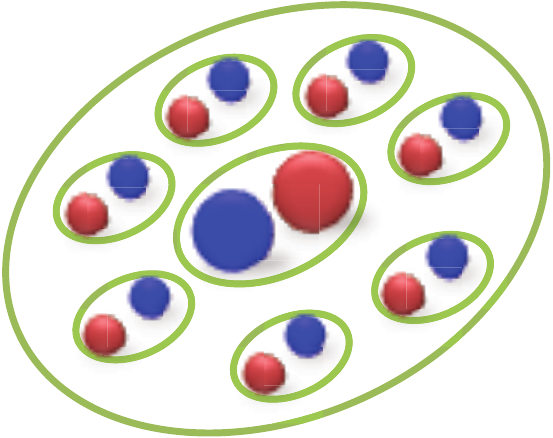} \hspace{1.0cm}\includegraphics[width=0.19\textwidth]{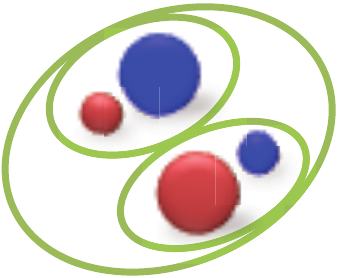}
\end{center}
\caption{Various scenarios of the exotic candidates, i.e. hybrid, tetraquark, hadroquarkonium and hadronic molecule (in order).
 The blue and red colors are used for quarks and antiquarks. Their size indicates the masses of the (anti-)quarks.}
\label{fig-1}
\end{figure}

The number of exotic candidates which are beyond the conventional quark model has grown dramatically during the last decades. 
Their appearance gives us challenges for understanding of the strong interaction.
Rencently, numerious review papers~\cite{Chen:2016qju,Esposito:2016noz,Lebed:2016hpi,Guo:2017jvc,Olsen:2017bmm}
  are written focusing on different properties. 
From the theoretical side, various scenarios, such as hybrid, tetraquark, hadroquarkonium and hadronic molecule
as shown in Fig.~\ref{fig-1}, are proposed. A hybrid is a compact heavy $Q\bar{Q}$ with excited gluons.
 A tetraquark is a compact object formed by a diquark and an anti-diquark. Hadroquarkonium is a compact $Q\bar{Q}$ 
 embedded in light quark cloud. On the contrary, a hadronic molecule is an extended object made of two or more hadrons. 
 The idea of a hadronic molecule is based on the observation that the deuteron exists as a shallow bound state of a proton and a neutron.
  It is also based on the fact that most (not all) of the exotic candidates are close to some $S$-wave thresholds with 
  narrow constituents, e.g. the two $Z_b$ states have masses close to the $B\bar{B}^*$ and the $B^*\bar{B}^*$ thresholds, respectively,
   the two $Z_c$ states have masses close to the $D\bar{D}^*$ and the $D^*\bar{D}^*$ thresholds, respectively, 
   the $X(3872)$  has a mass close to the 
   $D\bar{D}^*$ threshold and the $Y(4260)$ has a mass 
   close to the $D_1\bar{D}$ threshold. Another interesting feature is that
   splittings are predominantly given by those of the relevant nearby thresholds, e.g.
\begin{eqnarray}
m_{Y(4260)}-m_{X(3872)}\simeq m_{D_1}-m_{D^*},\quad m_{Z_c(4020)}-m_{Z_c(3900)}\simeq m_{D^*}-m_D.
\end{eqnarray}
This suggests that they could be viewed as hadronic molecules. However, in principle, 
 all configurations with the same quantum numbers can mix with each other. 
 Here, we present some observables that should allow us to distinguish the extended molecular scenario 
 from other more compact scenarios.

\section{Identify hadronic molecules}
The size of a hadronic molecule $R$ is characterized by the inverse of its binding momentum $1/\gamma$, where $\gamma=\sqrt{2\mu E_B}$, 
with $\mu$ the reduced mass and $E_B$ the binding energy. For near-threshold states, the size is therefore
much larger than that of normal
 compact states, which allows us to deduce some physical quantities that are sensitive to their internal structure. 
 In the following subsections, we present some of them.
\subsection{Effective couplings} 
Decades ago, S.~Weinberg proposed a criterion~\cite{Weinberg:1965zz} which related the effective coupling
\begin{eqnarray}
g_{eff}^2=Zg_0^2=\frac{2\pi \gamma}{\mu^2}(1-Z)
\end{eqnarray}
to $1-Z$ the probability for finding the hadronic molecule component in the wave function,
with $Z$ the wave function renormalization constant and $g_0$ the bare coupling~\cite{Guo:2017jvc}. 
Here, $E_B$ and $\gamma=\sqrt{2\mu E_B}$ are the binding energy and binding momentum, respectively.  
For a pure hadronic molecule, $Z=0$ which produces the maximal
  effective coupling $g_{eff}$.
   As the branching ratio of $Z_b\to B\bar{B}^*+c.c.$ is about $85.6\%$ in spite of a significant phase space suppression,
    it indicates that the $Z_b$ could be dominated by the $B\bar{B}^*+c.c.$ hadronic molecule. 
    For a pure compact object, $Z=1$ which produces a vanishing effective coupling.
\subsection{Scattering length}
The $T$-matrix of the two-body continuum channel is~\cite{Guo:2017jvc}
\begin{eqnarray}
T_{NR}(E)=\frac{g_0^2}{E+E_B+g_0^2\mu/(2\pi)(ik+\gamma)}
\label{eq-1}
\end{eqnarray}
with the relative momentum $k=\sqrt{2\mu E}$.
On the other hand, one can match Eq.~(\ref{eq-1}) to the effective range expansion~\cite{Guo:2017jvc}
\begin{eqnarray}
T_{NR}=-\frac{2\pi}{\mu}\frac{1}{1/a+rk^2/2-ik}
\label{eq-2}
\end{eqnarray}
and obtain the scattering length and effective range as
\begin{eqnarray}
a=-2\frac{1-Z}{2-Z}\frac{1}{\gamma}+\mathcal{O}(1/\beta),\quad r=-\frac{Z}{1-Z}\frac{1}{\gamma}+\mathcal{O}(1/\beta).
\end{eqnarray}
For a pure hadronic molecule $Z=0$, $a=-1/\gamma$ and $r=0$. For a pure compact object $Z=1$, $a=0$ and $r=-\infty$. 
The scattering length of isospin singlet $DK$ scattering from a Lattice calculation~\cite{Mohler:2013rwa} is $a_{DK}^{I=0}=-(1.33\pm 0.20)~\mathrm{fm}$. 
That value is consistent with the estimation $1/\gamma=1.05\pm0.25~\mathrm{fm}$ based on the pure $DK$ hadronic molecule scenario. The uncertainty of $0.25~\mathrm{fm}$ is from the higher order contribution $\mathcal{O}(1/\beta)$ with $\beta$ the mass of the exchanged $\rho$. 
The above estimation indicates that the $D_{s0}^*(2317)$ could be an isospin singlet $DK$ hadronic molecular state.
\subsection{Pole counting approach}
The zeros of the denominator of Eq.~(\ref{eq-2}) correspond to the poles of the scattering amplitude. In the $k$-plane, 
the two poles are located at~\cite{Baru:2003qq}
\begin{eqnarray}
k_1=i\gamma,\quad k_2=-i\gamma \left(\frac{2-Z}{Z}\right).
\end{eqnarray}
For a pure hadronic molecule $Z=0$, only the first pole is located close to the threshold with the second pole approaching negative infinity along the imaginary axis. For a pure compact  object $Z=1$, the two poles $k_1=i\gamma$ and $k_2=-i\gamma$ are distributed symmetrically respective to the corresponding threshold. Thus the number of poles near threshold could also help to distinguish the extended scenario from the compact ones~\cite{Morgan:1992ge}. 
\subsection{Line shapes in inelastic channels}
\begin{figure}
\begin{center}
\includegraphics[width=0.45\textwidth]{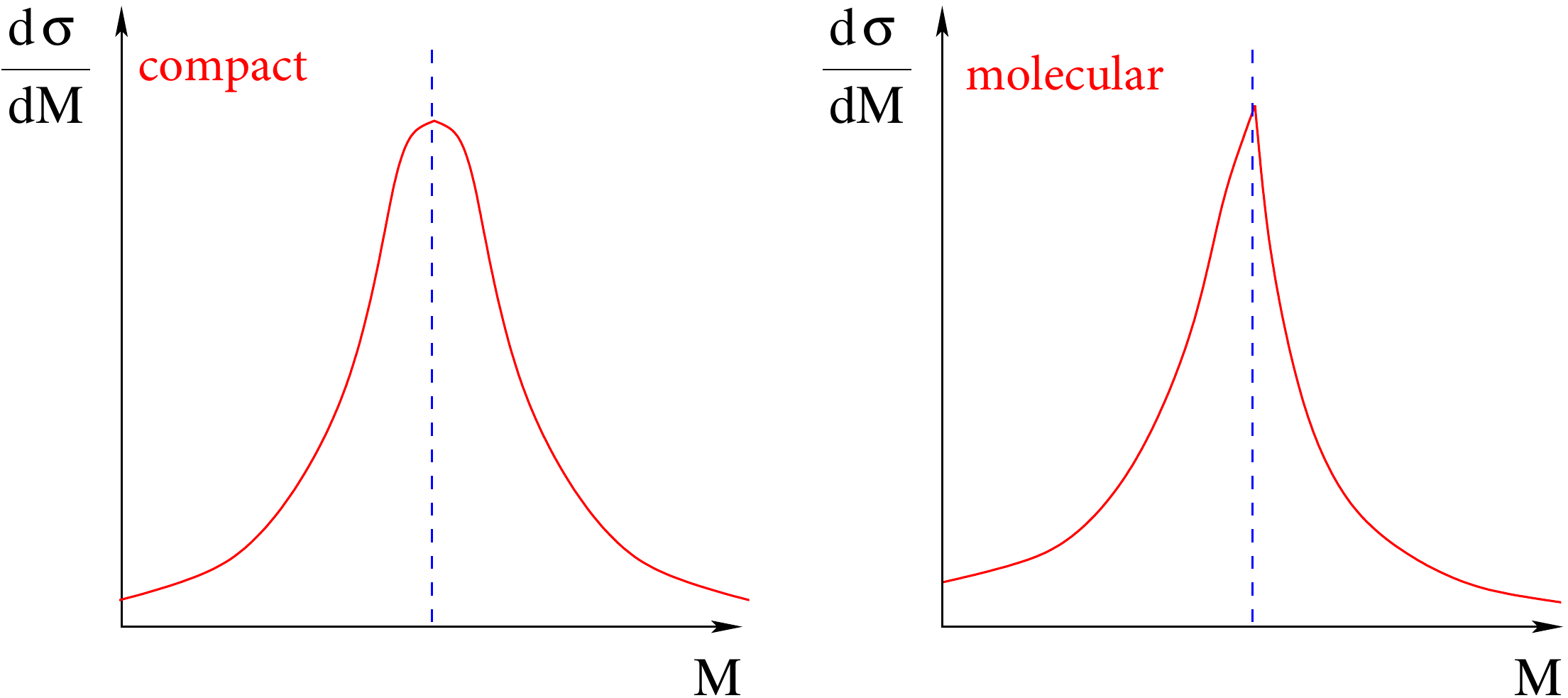}\includegraphics[width=0.28\textwidth]{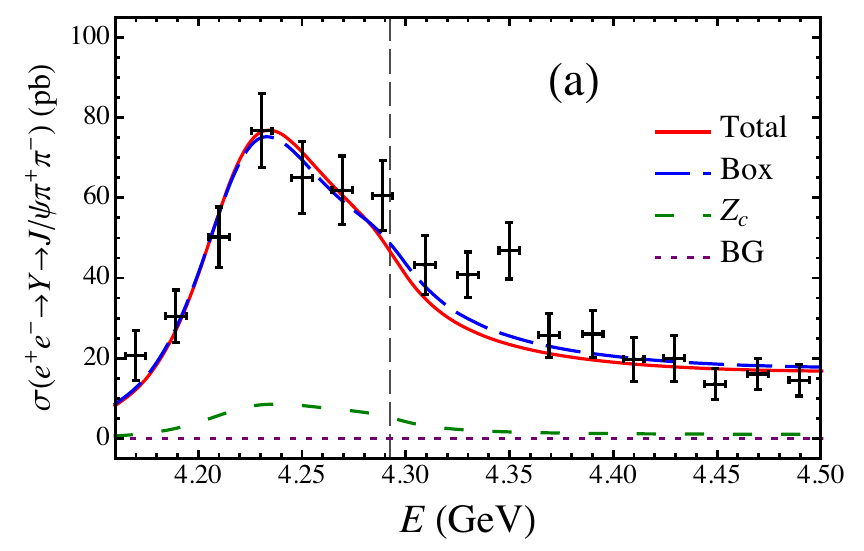}\includegraphics[width=0.28\textwidth]{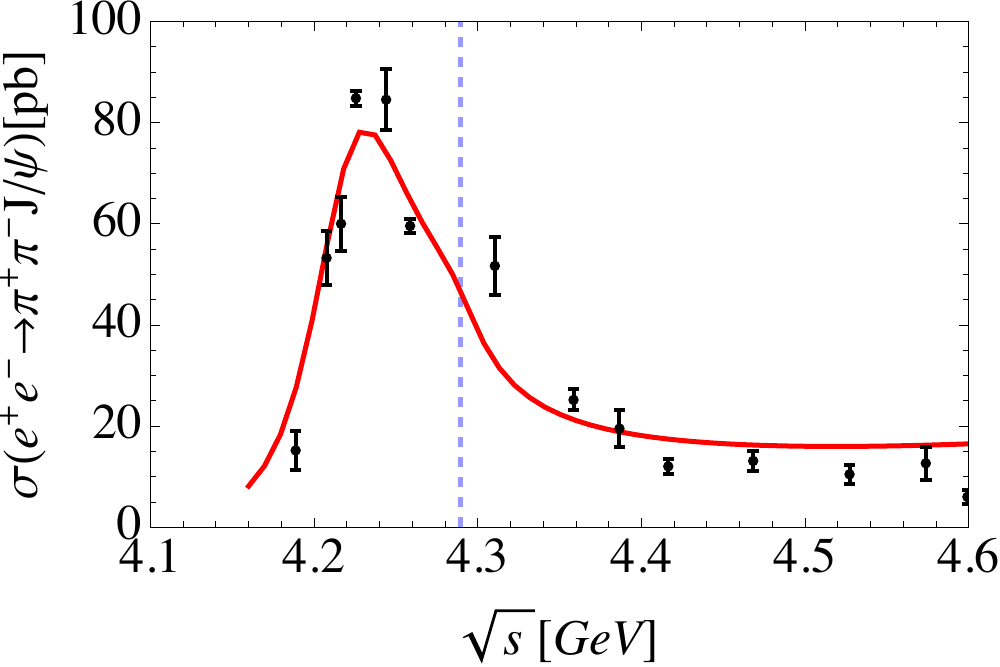}
\end{center}
\caption{The line shapes of inelastic channels near threshold for compact (the first figure) and molecular (the second one) state. The vertical lines 
are the relevant thresholds. The third figure is the fitted line shape of the $Y(4260)$~\cite{Cleven:2013mka} in $J/\psi\pi\pi$ channel with data from Belle~\cite{Yuan:2007sj}. The last figure is the fitted line shape in the third figure comparing to the recent data from BESIII~\cite{Ablikim:2016qzw}.} 
\label{fig-2}
\end{figure}

All the exotic candidates are unstable and have been measured in various channels. Thus the above scattering amplitude needs some 
modifications as~\cite{Guo:2017jvc}
\begin{eqnarray}
T_{in}\propto \frac{\sqrt{\Gamma_0}}{E-E_r+(g_{eff}^2)/2(ik+\gamma)+i\Gamma_0/2}
\end{eqnarray}
with the energy $E=k^2/(2\mu)$ and $\Gamma_0$ the partial width to the inelastic channels 
which enter the interaction perturbatively. 
For a pure compact scenario, the small effective coupling $g_{eff}$ means that the first term in the denominator plays an
 important role for the line shapes in inelastic channels. Thus the corresponding line shapes are symmetric as shown by 
 the first figure of Fig.~\ref{fig-2}. Alternatively, the large effective coupling in molecular picture means that the third term 
 dominates the behaviour of the line shapes in inelastic channels, leading to antisymmetric line shapes as shown by the
  second figure of Fig.~\ref{fig-2}. A typical antisymmetric example is the line shape of the $Y(4260)$ in the $J/\psi\pi\pi$ 
  channel in Fig.~\ref{fig-2}. The third figure
   is the fitted~\cite{Cleven:2013mka} line shape of the $Y(4260)$ based on the $D_1\bar{D}$ hadronic molecular picture,
    comparing to Belle data~\cite{Yuan:2007sj}, from which one can see a clear antisymmetric behaviour. 
    The fitted result also agrees with the recent data from BESIII~\cite{Ablikim:2016qzw}. From these line shapes, one can conclude 
   that the $Y(4260)$ could be a the $D_1\bar{D}$ molecular candidate.

\section{The line shapes of the two $Z_b$ stastes}
In the next step, we use the line shapes of the two $Z_b$ states as an example to extract the physical quantities which are sensitive to their internal structure. As we know that the Breit-Wigner parametrization only works for an isolated narrow state and the sum of them violates unitarity,  we proposed a practical parametrization scheme for these near-threshold states~\cite{Hanhart:2015cua,Guo:2016bjq}, which satisfies all the requirements of the $S$-matrix, such as unitarity and analyticity. It is also easy to be implemented in experimental analyses. The formulae are deduced based on some assumptions which will not affect the final conclusion for the current exotic near-threshold states. As most of the exotic candidates in the heavy quarkonium sector are observed in channels with one heavy quarkonium and some pions, they at least have a pair of heavy quarks. The small scattering length of the scattering between heavy quarkonium and pion on the lattice studies~\cite{Liu:2008rza,Detmold:2012pi} indicates that their interaction is weak. As the result,  we set them to zero. Furthermore we neglect the possible left hand cuts. The last assumption is that we work with separable interactions. In the end, we will illustrate that the inclusion of the non-separable interaction, such as One-Pion-Exchange (OPE) potential, will not change the final conclusion. The details of this practical parametrization can be found in Refs.~\cite{Hanhart:2015cua,Guo:2016bjq}. 

\begin{figure}
\begin{center}
\includegraphics[width=0.3\linewidth]{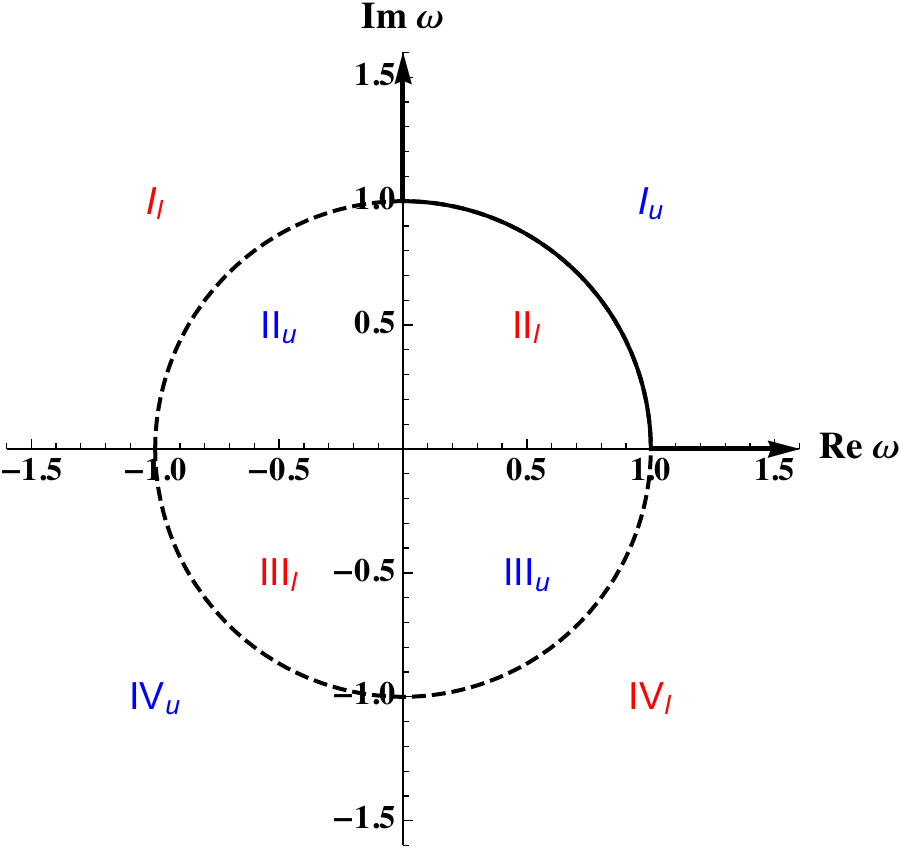}\includegraphics[width=0.30\linewidth]{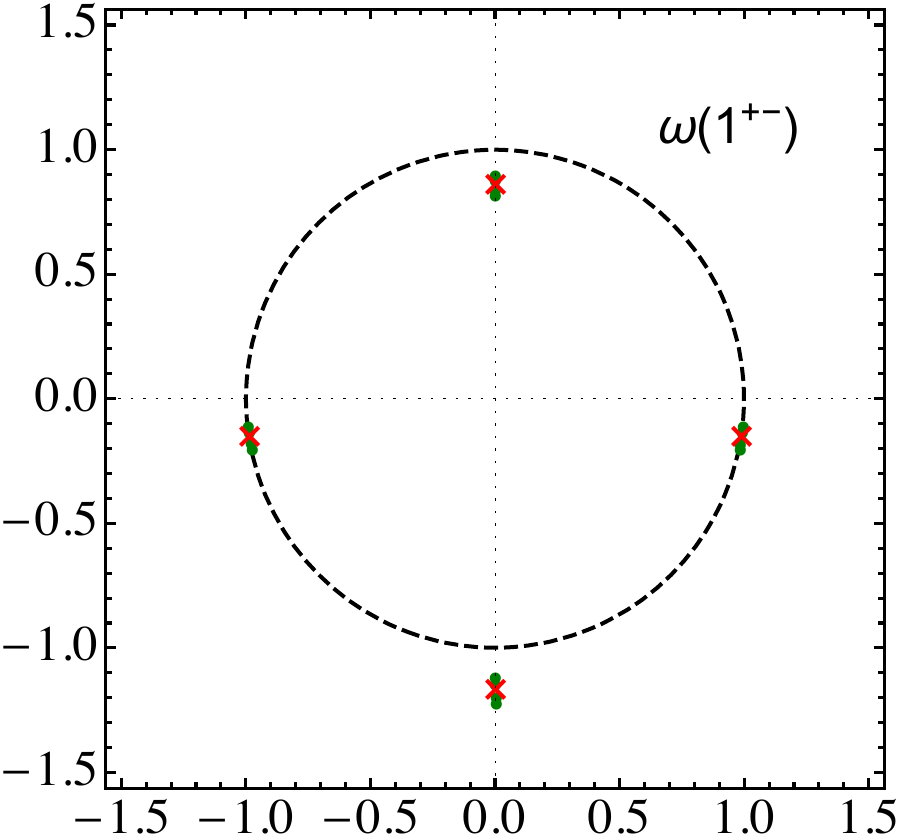} \includegraphics[width=0.30\linewidth]{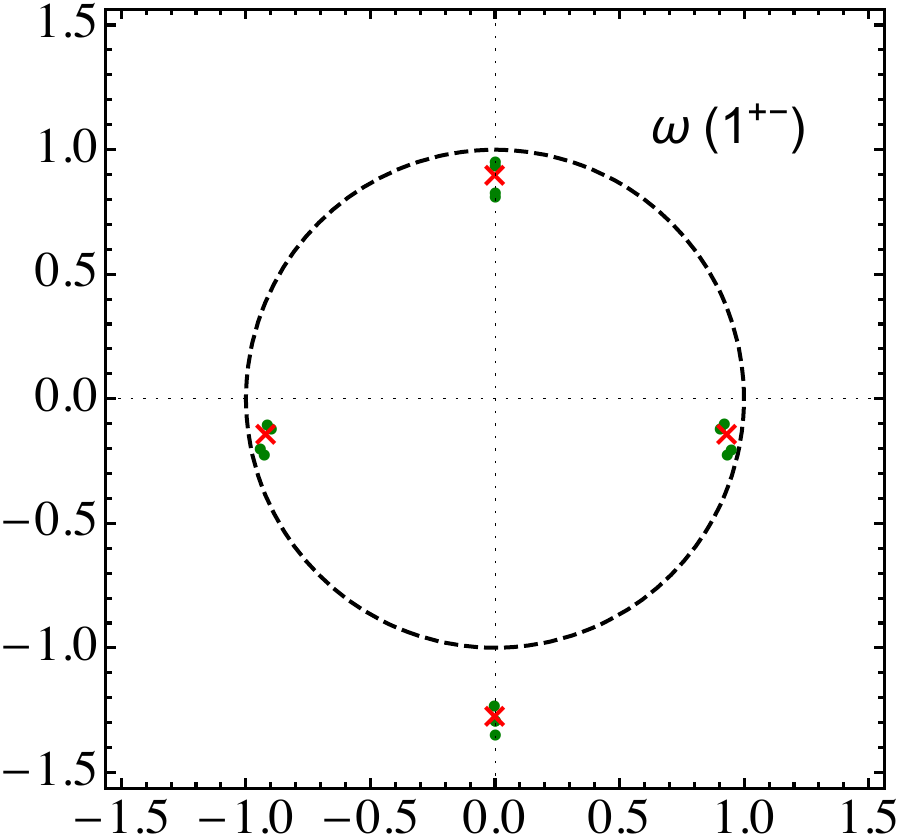}
\end{center}
\caption{First plot: The unitary-cut-free complex $\omega$-plane for the two elastic channels, i.e. $B\bar{B}^*$ and $B^*\bar{B}^*$.
 The subindices $u$ and $l$ mean the corresponding upper and lower half energy plane. The thick solid curve indicates the physical 
 region. Second plot: the pole positions on the $\omega$-plane for the fit respecting HQSS. 
 Third plot: the pole positions on the $\omega$-plane for the fit allowing sizeable HQSS breaking effect.}
\label{fig-3}
\end{figure}

As these two $Z_b$ states have a pair of bottom quarks, we expect that heavy quark spin symmetry (HQSS) works well here. 
Thus we produce the potential among elastic channels and inelastic channels which respects HQSS with additional parameters
 to control the HQSS breaking effect. We perform two fit schemes. 
 One respects HQSS and another one allows for HQSS breaking.
  These two fit schemes can describe the experimental 
  data equally well. With the fitted parameters, we search for poles of the two $Z_b$ states on the $\omega$ plane
   defined by the conformal transformation $k_1=\sqrt{\frac{\mu_1\delta}{2}}\left(\omega+\frac{1}{\omega}\right)$ 
   and $k_2=\sqrt{\frac{\mu_2\delta}{2}}\left(\omega-\frac{1}{\omega}\right)$ from the $k$ plane to the $\omega$ plane.
   Here, $k_1$ and $k_2$ are the three momenta of the $B\bar{B}^*$ and the $B^*\bar{B}^*$ channels in the center-of-mass frame, respectively.
$\mu_i$ is the reduced mass of the $i$ th channel and $\delta=m_{B^*}-m_B$ is the energy gap of these two elastic channels.
The advantage of the $\omega$ plane is that it is free of unitarity cut. 
   Although, it is the seven channel problem, one can search for poles on the Riemann sheet corresponding
 to the two elastic channels, since the contribution from the inelastic channels are marginal. 
 The classification of Riemann sheets are as below
\begin{eqnarray}
\mathrm{RS-I}:&&\quad{\rm Im}~k_1>0,\quad{\rm Im}~k_2>0,\quad\quad \mathrm{RS-II}:\quad{\rm Im}~k_1<0,\quad{\rm Im}~k_2>0,\\
\mathrm{RS-III}:&&\quad{\rm Im}~k_1>0,\quad{\rm Im}~k_2<0,\quad\quad \mathrm{RS-IV}:\quad{\rm Im}~k_1<0,\quad{\rm Im}~k_2<0.
\end{eqnarray}
 The energy relative to the lower $B\bar{B}^*$ threshold is
\begin{eqnarray}
E=\frac{k_1^2}{2\mu_1}=\frac{k_2^2}{2\mu_2}+\delta=\frac{\delta}{4}\left(\omega^2+\frac{1}{\omega^2}+2\right).
\end{eqnarray}
The different Riemann sheets in the $\omega$-plane are labeled as shown in the first figure of Fig.~\ref{fig-3}. The second and third figure of Fig.~\ref{fig-3} show the pole location of the two fit schemes. The upper and right-hand side poles are close to the physical region and will have large impact on the observables. The binding energies of the two $Z_b$ states within the two fit schemes are around $1\mathrm{MeV}$ as shown in Table~\ref{tab-1}. In the HQSS limit, the fit gives $\varepsilon_B(Z_b)=\varepsilon_B(Z_b^\prime)$ which is a consequence of the so-called Light Quark Spin Symmetry (LQSS) as proposed by Voloshin~\cite{Voloshin:2016cgm}. In the HQSS breaking fit, the two binding energies have some deviations.

\begin{figure}
\begin{center}
\includegraphics[width=0.99\textwidth]{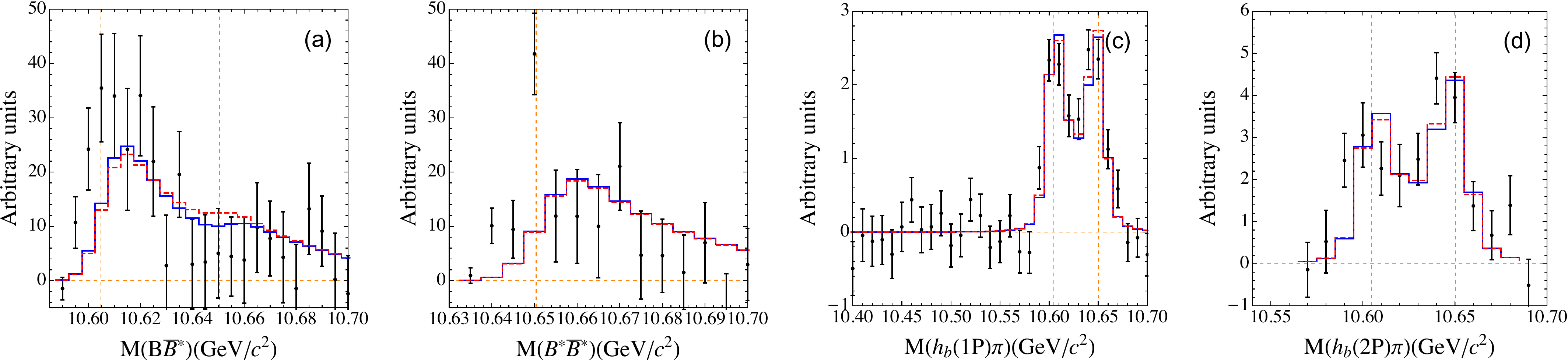} 
\end{center}
\caption{The line shapes of the two $Z_b$ states in the two elastic channels and $h_b(mP)\pi$
 with $m=1,2$ channels. The blue solid and red dashed curves are without and with non-separable
  OPE potential, respectively.}
\label{fig:comparison}
\end{figure}

To further test the effect of the non-separable interaction, we perform a fit including the OPE potential.
 As shown in Fig.~\ref{fig:comparison}, the two fits~\cite{zb}, i.e. without and with OPE potential, can describe the experimental data equally well. 
It reflects the fact that the line shapes are mainly driven by the pole position. On the other hand, 
it also confirms the validity of the parametrization. Furthermore, we extract the effective couplings of the
 relevant channels and find that those of $Z_bB\bar{B}^*$ and $Z_b^\prime B^*\bar{B}^*$
  are at least one order larger than the others~\cite{zb}.  These two large effective couplings indicate that
  $Z_b$ and $Z_b^\prime$ are dominated by the $B\bar{B}^*$ and $B^*\bar{B}^*$ hadronic molecules, respectively.

\begin{table}
\begin{center}
\caption{The binding energies of the two $Z_b$ states with the two fit schemes.}
\begin{tabular}{ccc}
\hline\hline
$\mathrm{MeV}$ & HQSS limit & HQSS breaking\\
\hline
$\varepsilon_B(Z_b)$& $1.10_{-0.54}^{+0.79}\pm i 0.06_{-0.02}^{+0.02}$ & $0.60_{-0.49}^{+1.40}\pm i 0.02_{-0.01}^{+0.02}$\\
$\varepsilon_B(Z_b')$ &  $1.10^{+0.79}_{-0.53} \pm i 0.08_{-0.05}^{+0.03}$ &$0.97^{+1.42}_{-0.68} \pm i 0.84_{-0.34}^{+0.22}$\\
\hline\hline
\end{tabular}
\end{center}
\label{tab-1}
\end{table}

\section{Summary and Outlook}
The measurement of exotic candidates challenges us understanding of the strong interaction.
 In principle, all the configurations with
the same quantum numbers can mix with each other. Thus at present we aim at identifying the prominent component. 
In this proceeding, we provide some quantities which can help to distinguish the extended hadronic molecule
 from more other compact scenarios for the near-threshold states.
To further pin down the nature of these exotic candidates, efforts are needed from both experimental and theoretical sides. 
From the experimental side, further measurements with high accuracy, such as partial widths, 
line shapes, quantum numbers and so on are needed. In addition, there should be searches for
 states in different channels. From the theoretical side, 
  a comparison with some results from the lattice, such as scattering length and spectrum,
  could help to clarify the situation. Furthermore, more observables should be identified that are
  sensitive to the internal structure of the exotics. 
\section{Acknowledgments}
The results presented here are based on the various collaborations with 
 Martin Cleven, Feng-Kun~Guo, Christoph~Hanhart, Yulia~Kalashnikova,  Patrick~Matuschek, Ulf-G.~Mei{\ss}ner, Roman~Mizuk,
 Alexey~Nefediev, Jan-Lukas~Wynen, Qiang~Zhao and Bing-Song~Zou, which are very appreciated by the author.
 This work is supported in part by the DFG (Grant No. TR110) and the NSFC (Grant No. 11621131001) through funds 
provided to the Sino-German CRC 110 ``Symmetries and the Emergence of Structure
in QCD''.

\end{document}